\documentclass[preprint]{elsarticle}
\usepackage{bm}
\usepackage[T2A]{fontenc}
\usepackage[cp1251]{inputenc}
\usepackage[english]{babel}
\usepackage{amsfonts,amssymb,amsmath}
\usepackage{amsmath,graphicx}
\usepackage{graphicx,pifont} 
\usepackage{epsfig}
\usepackage{parallel}
\title{Relativistic kinetic theory of classical systems of charged particles: towards the microscopic foundation of thermodynamics and kinetics}
\author[ayz]{A.Yu.~Zakharov} 
\fntext[fn1]{The work is supported by the Ministry of Science and Higher Education of Russian Federation within the framework of the project part of the state order (project No.~3.3572.2017, 2017--2019).}
\address{ Yaroslav the Wise Novgorod State University \\ Veliky Novgorod, 173003, Russia}
\begin{document}
	\begin{abstract}

\begin{center}
	Highlights
\end{center}
\begin{itemize}
	\item Microscopic classical dynamics of a system of charges with excluded field variables
	\item Retarded interactions between particles instead of the molecular chaos hypothesis
	\item Irreversibility of dynamics as a common property of many-body and few-body systems
	\item Microscopic model of heredity of many-body systems
	\item Non Laplacean determinism
\end{itemize}

In the complete system of equations of evolution of the classical system of charges and the electromagnetic field generated by them, the field variables are excluded. An exact closed relativistic non-Hamiltonian system of nonlocal kinetic equations, that describes the evolution of a system of charges in terms of their microscopic distribution functions, 
is obtained . The solutions of this system of equations are non-invariant with respect to time reversal, and also have the property of hereditarity.

	\end{abstract}
	
	\begin{keyword}
	Relativistic kinetics; irreversibility; many-body and few-body systems; retarded interactions; hereditarity phenomenon; non-Hamiltonian dynamics 
		
	\end{keyword}
	

\maketitle	


\section{Introduction}

%
%

Currently, theoretical studies of the thermodynamic and kinetic properties of condensed systems, gases, and plasma use mainly the ideas of equilibrium and non-equilibrium statistical mechanics. Classical statistical mechanics is based on two core principles.

\begin{itemize}
	\item The system of interacting particles is characterized by its Hamiltonian $ H\left(p_{i}(t), q_{i}(t) \right)  $, depending on the generalized coordinates $ q_ {i} (t) $ and momenta $ p_ {i } (t) $, $ i = 1,2, \ldots N $ ($ N $~is the number of degrees of freedom of the system). The evolution of this system in time is described by the Hamilton system of equations.	
	\item Probabilistic measures (microcanonical, canonical and grand canonical distributions) are defined in the phase space of the system. The choice of probabilistic measures depends on the external conditions imposed on the system.
\end{itemize}

Thus, in the framework of statistical mechanics, the thermodynamics of a many-particle system is uniquely determined by its Hamiltonian, i.e. has a purely mechanical origin.

The calculation of the thermodynamic functions of systems is reduced to solving of three problems.

\begin{itemize}
	\item 	The choice of inter-particle potentials. It is usually assumed that this problem is outside the scope of statistical mechanics. As interatomic potentials, some model potentials are used, chosen from considerations of simplicity and more or less feasible physical reasons.
	\item The choice of a probabilistic measure in the phase space of the system corresponding to the external conditions imposed on the system. For example, for a given temperature $ T $, volume $ V $ and the number of particles $ N $ of the system, the canonical distribution is used.
	
	\item Calculation of partition functions of systems in the thermodynamic limit, or uncoupling of BBGKY chains with the inevitable participation of additional hypotheses such as the principle of correlations weakening, etc.
	
\end{itemize}

It should be noted that the solution to the third of these problems meets enormous mathematical difficulties, which were got over only for a few extremely simplified models, very remote from ``reality''~ \cite{Baxter}.

However, besides to the purely mathematical problems of constructing the microscopic thermodynamics of many-body systems, there are fundamental questions, the answers to which can unlikely be found in the framework of statistical mechanics.

\begin{enumerate}
	\item What is the nature of thermodynamic equilibrium? What is the difference between the equilibrium and non-equilibrium states of a system in the framework of its microscopic dynamics? What is the real physical mechanism \textbf{irreversibly} leading an initially nonequilibrium isolated system to a state of thermodynamic equilibrium? Within the framework of the statistical approach \textbf {it is assumed} that the equilibrium state is the most probable state. However, this assumption is nothing more than a hypothesis accepted as an axiom~\cite{Kubo1, Kubo2}.

	\item Is it possible a consistent combination of the concepts of deterministic classical mechanics with probabilistic concepts introduced into the kinetic theory of matter by Maxwell, Boltzmann and Gibbs? Note that at the end of the 19th century not only atomism, but also Boltzmann's probabilistic approach was severy criticized by a group of physicists and mathematicians (Mach, Duhem, Poincar\'{e}?, Kirchhoff, Ostwald, Helm, etc.)~\cite {Brush3, Guttmann}. In fact, the Loschmidt and Zermelo paradoxes that have not found conclusive solution are internal contradictions in Boltzmann's kinetic theory~\cite {Strien,Shenker2}.
	
	Nevertheless, at present, most of the efforts of researchers are aimed at developing probabilistic methods in the theory of many-body problems. It is assumed that the abundance of the number of degrees of freedom of many-body systems and the instability of phase trajectories of Hamiltonian systems (dynamic chaos)~\cite{Zaslavsky, Hirsch, Argyris} are sufficient foundation for applying probabilistic concepts in classical mechanics.

	\item  Theoretically, the situation changed radically after the work of Kac~\cite{Kac, Gottwald}, in which a quite convincing counterexample to statistical mechanics was found, namely, an exactly solved dynamic ring model.  It is shown that the irreversible behavior of this dynamic model is a consequence of the incorrect use of seemingly feasible probabilistic hypotheses such as the assumption of molecular chaos. Despite this, the concept of probability continues to be used as an apparatus in statistical mechanics, primarily due to the lack of a constructive alternative concept.

	\item Classical mechanics is in obvious contradiction with the thermodynamic behavior of real systems, which raises quite reasonable doubts about the fundamental possibility of microscopic foundation of thermodynamics in the framework of classical mechanics~\cite {Brush3,Guttmann, Strien, Shenker2}. Therefore, it remains unclear
	are such fundamental problems as the correct explanation of the mechanisms leading to phase transitions, a consistent explanation of the nature of the second law of thermodynamics, and its reducibility (or irreducibility) to the more fundamental laws of physics solvable within the framework of existing theories~\cite{Lieb, Lieb1} .

\end{enumerate} 

Thus, statistical mechanics does not reveal the real physical mechanism of the thermodynamic behavior of real many-body systems, and existing attempts to explain the irreversibility using the concept of probability in combination with classical Hamiltonian dynamics seem to be exceptionable.

In addition, in the framework of statistical mechanics, the question of describing the interactions between the structural units of a substance (atoms, molecules, ions, free radicals, and others) remains open. In particular, even the idea of the existence of these structural units is rather limited, especially in the case of such species of condensed matter, as soft matter, disordered systems, electrolytes, etc. Finally, even the provisory distinguished structural units of matter~are not some ``rigid'' structures with a fixed structures, but complex systems with interacting internal degrees of freedom.

It is known that all ``real'' interactions between particles of matter are of electromagnetic origin. Therefore, strictly speaking, instead of a system of ``hard'' particles with model fixed interactions between them, an extended system consisting of charged particles and the electromagnetic field generated by these particles should be considered. 

In this regard, it is of interest to study the classical dynamics of a closed system of point charges interacting through the electromagnetic field created by them. Such a system consists of two subsystems: charges and an electromagnetic field. The evolution of this extended system is described by the Maxwell equations for the electromagnetic field and the equations of dynamics of charged particles.

The purpose of this paper is as follows.
\begin{enumerate}
	\item Construction of the classical microscopic kinetic theory of systems consisting of classical charged point particles and the electromagnetic field created by them. For this, it is necessary to exclude field variables in the dynamics equations of this system.
	\item Derivation of a complete system of equations describing the dynamics of particles with excluded field variables. Qualitative analysis of the properties of solutions of the equations of evolution for a system of interacting charges.
\end{enumerate}

\section{Microscopic distribution functions and equations of their evolution}

Consider an \ textbf{isolated classical} system consisting of a finite number of point charges interacting with each other through an electromagnetic field. The temporal evolution of this system as a whole is described by the coupled equations of motion of particles and Maxwell's equations. Of interest is the dynamics of the particle subsystem.

\subsection{Microscopic distribution functions}

We consider the dynamics of a system of particles in terms of microscopic distribution functions
\begin{equation}\label{f-alpha}
\begin{array}{r}
{\displaystyle 
	f_{\alpha}\left( \mathbf{r,p},t\right) = \sum_{s=1}^{N_{\alpha}}\, \delta\left(\mathbf{r}-\mathbf{R}_{s}^{\alpha}\left(t \right)  \right)\,  \delta\left(\mathbf{p}-\mathbf{{P}}_{s}^{\alpha}\left(t \right)  \right)  }\\ {\displaystyle =\iint \frac{d\mathbf{k}\, d\mathbf{q}}{\left( 2\pi \right)^6}\, e^{i\,\mathbf{k\cdot r} + i\, \mathbf{q\cdot p}}\, \sum_{s=1}^{N_{\alpha}}\, \left[e^{-i\,\mathbf{k\cdot R}_{s}^{\alpha}\left(t \right) }\, e^{-i\,\mathbf{q\cdot P}_{s}^{\alpha}\left(t \right) } \right], }
\end{array}
\end{equation}
where $ \mathbf{R}_{s}^{\alpha}\left(t \right) $ and $ \mathbf{P}_{s}^{\alpha}\left(t \right) $ are  the coordinates and momentum of the $ s $~-th particle at the time instant~$ t $, $ N_{\alpha} $~ is the total number of particles of the $ \alpha $~-th type. Note that by definition, microscopic distribution functions are not probabilistic functions.

Sums over $ s $ of arbitrary ``one-particle functions''  $ \psi_{\alpha}\left(\mathbf{R}_{\alpha}\left(t \right),  \mathbf{P}_{\alpha}\left(t \right) \right) $ are expressed through microscopic distribution functions using the identity
\begin{equation}\label{sum-s}
\sum_{s=1}^{N_{\alpha}}\, \psi_{\alpha}\left(\mathbf{R}_{\alpha}\left(t \right),  \mathbf{P}_{\alpha}\left(t \right) \right) = \iint d\mathbf{r}\, d\mathbf{p} \, f_{\alpha}\left( \mathbf{r,p},t\right) \, \psi_{\alpha}\left( \mathbf{r,p} \right). 
\end{equation}

In particular, for each type of particle their instantaneous microscopic particle densities $n_{\alpha}\left(\mathbf{r},t \right)  $, charge densities $ \rho_{\alpha}\left(\mathbf{r},t \right) $, flows  $ \mathbf{v}_{\alpha}\left(\mathbf{r},t \right) $ and electric currents $ \mathbf{j}_{\alpha}\left(\mathbf{r},t \right) $ in the system are functionals of microscopic distribution functions and have the following form:

\begin{equation}\label{n-alpha-r}
n_{\alpha}\left(\mathbf{r},t \right) =  \int d\mathbf{p}\, f_{\alpha}\left( \mathbf{r,p},t\right),
\end{equation}
\begin{equation}\label{rho-alpha-r}
\rho_{\alpha}\left(\mathbf{r},t \right) = Z_{\alpha}\, \int d\mathbf{p}\, f_{\alpha}\left( \mathbf{r,p},t\right).
\end{equation}
\begin{equation}\label{v-alpha-r}
\mathbf{v}_{\alpha}\left(\mathbf{r},t \right) = c\int d\mathbf{p}\, \frac{\mathbf{p}}{\sqrt{\left( p\right) ^{2}+m_{\alpha}^{2}c^{2} } } \, f_{\alpha}\left( \mathbf{r,p},t\right),
\end{equation}
\begin{equation}\label{j-alpha-r}
\mathbf{j}_{\alpha}\left(\mathbf{r},t \right) =  
Z_{\alpha}c\int d\mathbf{p}\, \frac{\mathbf{p}}{\sqrt{\left( p\right) ^{2}+m_{\alpha}^{2}c^{2} } } \, f_{\alpha}\left( \mathbf{r,p},t\right),
\end{equation}
where $ c $~is speed of light, $ p=\left| \mathbf{p}\right|  $, $ m_{\alpha} $~is mass of $ \alpha $-th type particle.

Here we have used the relation between the momentum of particle $ \mathbf{p}_{\alpha} $ and its velocity $ \mathbf{v}_{\alpha} $
\begin{equation}\label{v-P}
\mathbf{v}_{\alpha}\left(t \right) = \frac{\mathbf{p}_{\alpha}\left(t \right)\, c}{   \sqrt{\left( \mathbf{p}_{\alpha}\left(t \right) \right) ^{2}  + m_{\alpha}^{2}\, c^{2}}},
\end{equation}

\subsection{Kinematic and dynamic contributions to the evolution of microscopic distribution functions}

Differentiating the function~\eqref{f-alpha} with respect to time, we find
\begin{equation}\label{dot1+2-f}
\frac{\partial f_{\alpha}(\mathbf{r},\mathbf{p},t)}{\partial t} = \left( \frac{\partial f_{\alpha}(\mathbf{r},\mathbf{p},t)}{\partial t}\right) _{1} + \left( \frac{\partial f_{\alpha}(\mathbf{r},\mathbf{p},t)}{\partial t}\right) _{2},
\end{equation}
where 
\begin{equation}\label{dot1-f}
\begin{array}{r}
{\displaystyle \left( \frac{\partial f_{\alpha}(\mathbf{r},\mathbf{p},t)}{\partial t}\right) _{1} = \iint\, \frac{d\mathbf{k}\, d\mathbf{q}}{\left( 2\pi\right)^6 }\, e^{i\,\mathbf{k\cdot r} + i\,\mathbf{q\cdot p}} \times  }\\  {\displaystyle \times 
	\sum_{s=1}^{N_{\alpha}} \,e^{- i\, \mathbf{k\cdot R}_s^{\alpha} \left(t \right) } \,e^{-i\,\mathbf{q\cdot R}_s^{\alpha} \left(t \right) } \left( - i\, \mathbf{k\cdot \dot {R} }_s^{\alpha} \left(t \right)  \right) ,}
\end{array}
\end{equation}
and
\begin{equation}\label{dot2-f}
\begin{array}{r}
{\displaystyle \left( \frac{\partial f_{\alpha}(\mathbf{r},\mathbf{p},t)}{\partial t}\right) _{2} = \iint\, \frac{d\mathbf{k}\, d\mathbf{q}}{\left( 2\pi\right)^6 }\, e^{i\,\mathbf{k\cdot r} + i\,\mathbf{q\cdot p} }\times }\\ 
{\displaystyle \times  
	\sum_{s=1}^{N_{\alpha}} \,e^{- i\, \mathbf{k\cdot  R}_s^{\alpha} \left(t \right) } \,e^{-i\,\mathbf{q\cdot P}_s^{\alpha} \left(t \right) } \left( - i\, \mathbf{q \cdot \dot{P} }_s^{\alpha} \left(t \right)  \right) .}
\end{array}
\end{equation}
The quantities~\eqref{dot1-f} and~\eqref{dot2-f} are the kinematic and dynamic contributions to the rate of change of the microscopic distribution function of the particle system, respectively.

The expression for the kinematic contribution $ \left( \frac{\partial f_{\alpha}(\mathbf{r},\mathbf{p},t)}{\partial t}\right) _{1} $ using the identity~\eqref{sum-s}  is reduced to the form:
\begin{equation}\label{dot1-f2}
\begin{array}{r}
{\displaystyle \left( \frac{\partial f_{\alpha}(\mathbf{r},\mathbf{p},t)}{\partial t}\right) _{1} = \iint\, \frac{d\mathbf{k}\, d\mathbf{q}}{\left( 2\pi\right)^6 }\, e^{i\,\mathbf{k\cdot r} + i\,\mathbf{q\cdot p}}  }\\  {\displaystyle \times 
	\iint  \frac{c\ d\mathbf{r}'\, d\mathbf{p}'}{\sqrt{\left(p' \right)^{2} + m_{\alpha}^{2} c^{2}} } \, f_{\alpha}\left(\mathbf{r', p'}, t \right) \,e^{- i\, \mathbf{k\cdot r'} }  \,e^{-i\,\mathbf{q\cdot p'} }  \left( - i\, \mathbf{k\cdot p'} \right) }\\
{\displaystyle =  \frac{c}{\sqrt{p^{2} + m_{\alpha}^{2} c^{2}} } \int d\mathbf{r'} \left\lbrace  \int \frac{d\mathbf{k}}{\left(2\pi \right)^{3} }\,  e^{i \mathbf{k \cdot}  \left(\mathbf{ r-r'}\right) } \left( -i \mathbf{k \cdot p}\right)\right\rbrace f_{\alpha}\left(\mathbf{r', p}, t\right)   }\\
{\displaystyle  = - \frac{c}{\sqrt{p^{2} + m_{\alpha}^{2} c^{2}} } \left(\mathbf{p} \cdot \frac{\partial}{\partial \mathbf{r}} \right) f_{\alpha}\left(\mathbf{r, p}, t\right). }
\end{array}
\end{equation}

The situation with the calculation of $  \left( \frac{\partial f_{\alpha}(\mathbf{r},\mathbf{p},t)}{\partial t}\right) _{2} $ is somewhat more complicated, since here the equation of motion of particles needs to be used
\begin{equation}\label{dot-P}
\mathbf{\dot{P}}_{s}^{\alpha}\left( t\right)  = Z_{\alpha}\left[  \mathbf{E}\left( \mathbf{R}_{s}^{\alpha} \left(t \right), t \right) + \frac{1}{c}\, \mathbf{\dot{R}}_{s}^{\alpha} \times \mathbf{H}\left( \mathbf{R}_{s}^{\alpha} \left(t \right), t \right)\right],
\end{equation}
where $ \mathbf{E}\left( \mathbf{R}_{s}^{\alpha} \left(t \right), t \right) $ and $ \mathbf{H}\left( \mathbf{R}_{s}^{\alpha} \left(t \right), t \right) $~are strengths of electric and magnetic fields at a point $ \mathbf{R}_{s}^{\alpha} \left( t\right) $ at time $ t $. This equation expresses the instantaneous force $ \mathbf{\dot{P}}_{s}^{\alpha}\left( t\right)  $ acting on a particle as a function of its coordinates $ \mathbf{R}_{s}^{\alpha}\left( t\right)  $ and velocity $ \mathbf{\dot{R}}_{s}^{\alpha}\left( t\right)  $.  
 
Substituting~\eqref{dot-P} into~\eqref{dot2-f} allows us to reduce the sum the sum over $ s $ to the form needed for use the identity~\eqref{sum-s}. As a result, we have:

\begin{equation}\label{dot-P-EH}
\begin{array}{c}
{\displaystyle \sum_{s=1}^{N_{\alpha}} \,e^{- i\, \mathbf{k\cdot  R}_s^{\alpha} \left(t \right) } \,e^{-i\,\mathbf{q\cdot P}_s^{\alpha} \left(t \right) } \left( - i\, \mathbf{q \cdot \dot{P} }_s^{\alpha} \left(t \right)  \right)  }\\
{\displaystyle = Z_{\alpha} \iint d\mathbf{r'} d\mathbf{p'}\, f_{\alpha}\left(\mathbf{r', p'}, t \right) \,e^{- i\, \mathbf{k\cdot r'} }  \,e^{-i\,\mathbf{q\cdot p'} } }\\
{\displaystyle \times    \left( -i \mathbf{q\ \cdot }\left[\mathbf{E} \left(\mathbf{r'}, t \right)  +  \frac{\mathbf{p'} \times \mathbf{H}\left(\mathbf{r'}, t \right)}{\sqrt{\left(p' \right)^{2} + m_{\alpha}^{2}c^{2} } } \right]  \right) .}
\end{array}
\end{equation}
Substituting this expression into the formula~\eqref{dot2-f}, we obtain the following representation for the dynamic part of the rate of change of the microscopic distribution function:
\begin{equation}\label{dot2-f2}
\begin{array}{r}
{\displaystyle \left( \frac{\partial f_{\alpha}(\mathbf{r},\mathbf{p},t)}{\partial t}\right) _{2} = \iint\, \frac{d\mathbf{k}\, d\mathbf{q}}{\left( 2\pi\right)^6 }\, e^{i\,\mathbf{k\cdot r} + i\,\mathbf{q\cdot p} } }\\ 
{\displaystyle \times  
	Z_{\alpha} \iint d\mathbf{r'} d\mathbf{p'}\, f_{\alpha}\left(\mathbf{r', p'}, t \right) \,e^{- i\, \mathbf{k\cdot r'} }  \,e^{-i\,\mathbf{q\cdot p'} } }\\
{\displaystyle \times    \left( -i \mathbf{q\ \cdot }\left[\mathbf{E} \left(\mathbf{r'}, t \right)  +  \frac{\mathbf{p'} \times \mathbf{H}\left(\mathbf{r'}, t \right)}{\sqrt{\left(p' \right)^{2} + m_{\alpha}^{2}c^{2} } } \right]  \right) .}
\end{array}
\end{equation}

In the absence of external fields, the electromagnetic field contained in this formula can be expressed through the microscopic distribution functions of the system.

\subsection{The relationship of electromagnetic fields with microscopic distribution functions}
As it is known~\cite{Landau}, scalar $ \varphi \left(\mathbf{r},t \right) $ and vector $ \mathbf{A} \left(\mathbf{r},t \right) $ potentials of the electromagnetic field at time instant $ t $ depends on the spatial distribution of the charges $ \rho \left(\mathbf{r'},t' \right)  $ and the currents $ \mathbf{j} \left(\mathbf{r'},t' \right)$ of the system at all previous time instants $ t' \leq t $ according to Lienard-Wiechert relations
\begin{equation}\label{Lienard}
\left\lbrace 
\begin{array}{r}
{\displaystyle  \varphi \left(\mathbf{r},t \right) = \int  d\mathbf{r}'\ \frac{1}{\left| \mathbf{r} - \mathbf{r}'  \right| }\ \rho\left(\mathbf{r}', {t-\frac{\left| \mathbf{r} - \mathbf{r}'  \right|}{c}}
	 \right), 
}\\
{\displaystyle \mathbf{A} \left(\mathbf{r},t \right) = \frac{1}{c} \int  d\mathbf{r}'\ \frac{1}{\left| \mathbf{r} - \mathbf{r}'  \right| }\ \mathbf{j} \left(\mathbf{r}', {t-\frac{\left| \mathbf{r} - \mathbf{r}'  \right|}{c} }
	\right).}
\end{array}
\right. 
\end{equation}

We find the total charge density and current density at the point $ \mathbf{r'} $ at time instant  $t' = t - \frac{\left| \mathbf{r-r'}\right| }{c} $ taking into account the formulas~\eqref{rho-alpha-r} and~\eqref{j-alpha-r}

\begin{equation}\label{rho-full-r-tau}
\rho\left(\mathbf{r}',t - \frac{\left| \mathbf{r} - \mathbf{r}'\right| }{c}\right) = \sum_{\gamma} Z_{\gamma}\, \int d\mathbf{p}'\, f_{\gamma} \left( \mathbf{r',p'},t - \frac{\left| \mathbf{r} - \mathbf{r}'\right| }{c} \right),
\end{equation}

\begin{equation}\label{j-full-r-tau}
\mathbf{j}\left(\mathbf{r}',t - \frac{\left| \mathbf{r} - \mathbf{r}'\right| }{c} \right) =  \sum_{\gamma}
Z_{\gamma}c\int d\mathbf{p'}\, \frac{\mathbf{p'}}{\sqrt{\left( p'\right) ^{2}+m_{\gamma}^{2}c^{2} } } \, f_{\gamma}\left( \mathbf{r',p'},t - \frac{\left| \mathbf{r} - \mathbf{r}'\right| }{c} \right).
\end{equation}
From here we find expressions relating the microscopic distribution functions of particles with the potentials of the electromagnetic field
\begin{equation}\label{varphi(r,t)}
\varphi\left(\mathbf{r},t \right) = \sum_{\gamma}\, Z_{\gamma}\iint d\mathbf{r}' d\mathbf{p}'\,\left\lbrace  \frac{f_{\gamma}\left( \mathbf{r',p'},t - \frac{\left| \mathbf{r} - \mathbf{r}'\right| }{c}\right)}{\left| \mathbf{r} - \mathbf{r}'\right|}\right\rbrace  ;
\end{equation}
\begin{equation}\label{A(r,t)}
\mathbf{A}\left(\mathbf{r},t \right) =  \sum_{\gamma}\, Z_{\gamma} \iint d\mathbf{r}'  d\mathbf{p}'\, \frac{\mathbf{p}'}{\sqrt{\left( p'\right) ^{2}+m_{\gamma}^{2}c^{2} } } \left\lbrace  \frac{f_{\gamma}\left( \mathbf{r',p'},t - \frac{\left| \mathbf{r} - \mathbf{r}'\right| }{c}\right)}{\left| \mathbf{r} - \mathbf{r}'\right|}\right\rbrace  .
\end{equation}

We pass from the potentials of the electromagnetic field to the strengths of the electric $ \mathbf{E}(\mathbf{r}',t) $ and magnetic $ \mathbf{H}(\mathbf{r}',t) $ fields. As a result, we obtain the following relationships expressing the strengths of the electric and magnetic fields in terms of microscopic particle distribution functions:
\begin{equation}\label{E(r,t)1}
\begin{array}{r}
{\displaystyle \mathbf{E}(\mathbf{r}',t) =  -\nabla \varphi\left(\mathbf{r}',t \right) - \frac{1}{c} \frac{ \partial\mathbf{A} \left(\mathbf{r}',t \right) }{\partial t}  }\\
{\displaystyle = -\sum_{\gamma} Z_{\gamma} \iint  \frac{d\mathbf{r''} d\mathbf{p''} }{\left|\mathbf{r' - r''} \right| } \left\lbrace \frac{\mathbf{p''}}{\sqrt{\left( p''\right) ^{2} + m_{\gamma}^{2} c^{2}} }  \frac{1}{c}\ \frac{\partial f_{\gamma}\left( \mathbf{r'',p''},t - \frac{\left| \mathbf{r}' - \mathbf{r}''\right| }{c}\right)}{\partial t}   \right\rbrace   }\\
{\displaystyle + \sum_{\gamma} Z_{\gamma} \iint  \frac{d\mathbf{r''} d\mathbf{p''} }{\left|\mathbf{r' - r''} \right| ^{2}}  \left( \mathbf{r' - r''} \right)  }\\
{\displaystyle  
	\times  \left\lbrace  \frac{ f_{\gamma}\left( \mathbf{r'',p''},t - \frac{\left| \mathbf{r}' - \mathbf{r}''\right| }{c}\right) }{\left| \mathbf{r' - r''} \right|}  + 
	\frac{1}{c}\ \frac{\partial f_{\gamma}\left( \mathbf{r'',p''},t - \frac{\left| \mathbf{r}' - \mathbf{r}''\right| }{c}\right)}{\partial t}   \right\rbrace,}
\end{array}
\end{equation} 
\begin{equation}\label{H(r,t)0}
\begin{array}{r}
{\displaystyle \mathbf{H}(\mathbf{r}',t) = \nabla_{\mathbf{r}'} \times \mathbf{A}(\mathbf{r}',t)   }\\  \\
{\displaystyle =  \sum_{\gamma}\, Z_{\gamma} \iint \mathbf{dr''} \mathbf{dp''}  \frac{1 }{\sqrt{\left( p''\right) ^{2} + m_{\gamma}^{2} c^{2} } }  }\\ 
{\displaystyle \times \left[ \nabla_{\mathbf{r}'} \times  \frac{\mathbf{p}''}{\left| \mathbf{r}' - \mathbf{r}''\right|}   \,  f_{\gamma}\left( \mathbf{r'',p''},t - \frac{\left| \mathbf{r}' - \mathbf{r}''\right| }{c}\right)\right]  }\\
{\displaystyle = \sum_{\gamma}\, Z_{\gamma} \iint \mathbf{dr''} \mathbf{dp''} \left[  \frac{\mathbf{p''} }{\sqrt{\left( p''\right) ^{2} + m_{\gamma}^{2} c^{2} } } \times \frac{\mathbf{r'-r''}}{\left|\mathbf{r'-r''}   \right|^{2} } \right] }\\
{\displaystyle \times \left(\frac{1}{\left|\mathbf{r'-r''} \right|} + \frac{1}{c} \frac{\partial}{\partial t} \right) f_{\gamma}\left( \mathbf{r'',p''},t - \frac{\left| \mathbf{r}' - \mathbf{r}''\right| }{c}\right).}
\end{array}
\end{equation}

These expressions we use to calculate the dynamic contribution to the rate of change of the microscopic distribution functions~\eqref{dot2-f2}.

\subsection{Dynamic contribution to the evolution of the distribution function}
Let us separate the electric and magnetic components of the dynamic contribution to the rate of change of the microscopic distribution function~\eqref{dot2-f2}
\begin{equation}\label{dot2-f3}
\left( \frac{\partial f_{\alpha}(\mathbf{r},\mathbf{p},t)}{\partial t}\right) _{2} = \left( \frac{\partial f_{\alpha}(\mathbf{r},\mathbf{p},t)}{\partial t}\right) _{2,\mathbf{E}} + \left( \frac{\partial f_{\alpha}(\mathbf{r},\mathbf{p},t)}{\partial t}\right) _{2,\mathbf{H}},
\end{equation}
where 
\begin{equation}\label{dot2-fE}
\begin{array}{r}
{\displaystyle \left( \frac{\partial f_{\alpha}(\mathbf{r},\mathbf{p},t)}{\partial t}\right) _{2,\mathbf{E}} = Z_{\alpha} \iint\, \frac{d\mathbf{k}\, d\mathbf{q}}{\left( 2\pi\right)^6 }\, e^{i\,\mathbf{k\cdot r} + i\,\mathbf{q\cdot p} } }\\ 
{\displaystyle \times 
	\iint d\mathbf{r'} \, d\mathbf{p'} 
	\,e^{- i\, \mathbf{k\cdot r'}- i \mathbf{q\cdot p'} } f_{\alpha}(\mathbf{r', p'},t ) \left( - i\, \mathbf{q\cdot E}\left( \mathbf{r'}, t \right)  \right) ,}
\end{array}
\end{equation}
and
\begin{equation}\label{dot2-fH}
\begin{array}{r}
{\displaystyle \left( \frac{\partial f_{\alpha}(\mathbf{r},\mathbf{p},t)}{\partial t}\right) _{2,\mathbf{H}} = Z_{\alpha}\iint\, \frac{d\mathbf{k}\, d\mathbf{q}}{\left( 2\pi\right)^6 }\, e^{i\,\mathbf{k\cdot r} + i\,\mathbf{q\cdot p} } }\\ 
{\displaystyle \times  
 \iint d\mathbf{r'} d\mathbf{p'} 	\,e^{- i\, \mathbf{k\cdot r'}- i \mathbf{q\cdot p'} } f_{\alpha}\left(\mathbf{r', p'}, t \right) 
	 \left( -i \mathbf{q\ \cdot }\left[ \frac{\mathbf{p'} \times \mathbf{H}\left(\mathbf{r'}, t \right)}{\sqrt{\left(p' \right)^{2} + m_{\alpha}^{2}c^{2} } } \right]  \right).}
\end{array}
\end{equation}

Substitutions the expression~\eqref{E(r,t)1} into~\eqref{dot2-fE} and~\eqref{H(r,t)0} into~\eqref{dot2-fH} after lengthy but rather elementary transformations lead to the following results
\begin{equation}\label{dot2-fE-2}
\begin{array}{r}
{\displaystyle  \left( \frac{\partial f_{\alpha}(\mathbf{r},\mathbf{p},t)}{\partial t}\right) _{2,\mathbf{E}} = Z_{\alpha} \sum_{\gamma} Z_{\gamma}\iint \frac{d\mathbf{r'}\, d\mathbf{p'} }{\left|\mathbf{r-r'}\right| }  }\\
{\displaystyle \times  \left(\frac{\mathbf{p'}}{  \sqrt{\left( p'\right)^{2}+ m_{\gamma}^{2} c^{2}}}\cdot \frac{\partial}{\partial \mathbf{p}} \right)  f_{\alpha}\left(\mathbf{r},\mathbf{p},t \right) 
	\frac{1}{c}\, \frac{\partial f_{\gamma}\left( \mathbf{r'},\mathbf{p'},t-\frac{\left|\mathbf{r-r'} \right| }{c} \right) }{\partial t} }\\
{\displaystyle - Z_{\alpha} \sum_{\gamma} Z_{\gamma}\iint \frac{d\mathbf{r'}\, d\mathbf{p'} }{\left|\mathbf{r-r'}\right|^{2}  } \left(\left(\mathbf{r-r'} \right)\cdot \frac{\partial}{\partial \mathbf{p}}  \right) f_{\alpha}(\mathbf{r},\mathbf{p},t) }\\
{\displaystyle  \times \left[ \frac{1}{\left|\mathbf{r-r'} \right| }+ \frac{1}{c}\, \frac{\partial}{\partial t}\right] f_{\gamma}\left( \mathbf{r'},\mathbf{p'},t-\frac{\left|\mathbf{r-r'} \right| }{c} \right)  }
\end{array}
\end{equation}
and 
\begin{equation}\label{dot2-fH-3}
\begin{array}{c}
{\displaystyle \left( \frac{\partial f_{\alpha}(\mathbf{r},\mathbf{p},t)}{\partial t}\right) _{2,\mathbf{H}} = Z_{\alpha} \sum_{\gamma} Z_{\gamma} \iint  \frac{d\mathbf{r'}\, d\mathbf{p'}}{\left|\mathbf{r-r'} \right|^{2} } }\\
{\displaystyle \times \left(\mathbf{\left[p \times  \left[\left(r-r' \right) \times p' \right]\, \right] \cdot }\frac{\partial}{\partial \mathbf{p}} \left( \frac{f_{\alpha}(\mathbf{r},\mathbf{p},t)}{\sqrt{p^{2}+m_{\alpha}^{2}c^{2}} }\right) \right) }\\
{\displaystyle \times    \left(\frac{1}{\left| \mathbf{r-r'}\right|} + \frac{1}{c}\, \frac{\partial}{\partial t}  \right)\frac{f_{\gamma}\left( \mathbf{r'},\mathbf{p'},t- \frac{\left| \mathbf{r-r'}\right|}{c} \right) }{\sqrt{\left(p'\right)^{2}+m_{\gamma}^{2}c^{2}} },   }\\
\end{array}
\end{equation}
respectively. 

\section{Complete system of relativistic kinetic equations}

The equations of dynamics of a system of particles interacting with each other through an electromagnetic field are obtained by substituting the formulas~\eqref{dot1-f2}, \eqref{dot2-fE} and~\eqref{dot2-fH} into the equation~\eqref{dot1+2-f} and have the form:
\begin{equation}\label{bas-eq}
\begin{array}{r}
{\displaystyle \frac{\partial f_{\alpha}(\mathbf{r},\mathbf{p},t)}{\partial t} +  \frac{c}{\sqrt{p^{2} + m_{\alpha}^{2} c^{2}} } \left(\mathbf{p} \cdot \frac{\partial}{\partial \mathbf{r}} \right) f_{\alpha}\left(\mathbf{r, p}, t\right)}\\ 
{\displaystyle - Z_{\alpha} \sum_{\gamma} Z_{\gamma}\iint \frac{d\mathbf{r'}\, d\mathbf{p'} }{\left|\mathbf{r-r'}\right| }  
	\left[  \left(\frac{\mathbf{p'}}{  \sqrt{\left( p'\right)^{2}+ m_{\gamma}^{2} c^{2}}}\cdot \frac{\partial}{\partial \mathbf{p}} \right)  f_{\alpha}\left(\mathbf{r},\mathbf{p},t \right) \right]}\\
{\displaystyle   
\times 	\frac{1}{c}\, \frac{\partial f_{\gamma}\left( \mathbf{r'},\mathbf{p'},t-\frac{\left|\mathbf{r-r'} \right| }{c} \right) }{\partial t} }\\
{\displaystyle +  Z_{\alpha} \sum_{\gamma} Z_{\gamma}\iint \frac{d\mathbf{r'}\, d\mathbf{p'} }{\left|\mathbf{r-r'}\right|^{2}  }\left[  \left(\left(\mathbf{r-r'} \right)\cdot \frac{\partial}{\partial \mathbf{p}}  \right) f_{\alpha}(\mathbf{r},\mathbf{p},t)\right]  }\\
{\displaystyle  \times \left[ \frac{1}{\left|\mathbf{r-r'} \right| }+ \frac{1}{c}\, \frac{\partial}{\partial t}\right] f_{\gamma}\left( \mathbf{r'},\mathbf{p'},t-\frac{\left|\mathbf{r-r'} \right| }{c} \right)  }\\
{\displaystyle - Z_{\alpha} \sum_{\gamma} Z_{\gamma} \iint  \frac{d\mathbf{r'}\, d\mathbf{p'}}{\left|\mathbf{r-r'} \right|^{2} } \left[  \left(\frac{1}{\left| \mathbf{r-r'}\right|} + \frac{1}{c}\, \frac{\partial}{\partial t}  \right)\frac{f_{\gamma}\left( \mathbf{r'},\mathbf{p'},t- \frac{\left| \mathbf{r-r'}\right|}{c} \right) }{\sqrt{\left(p'\right)^{2}+m_{\gamma}^{2}c^{2}} } \right]  }\\
{\displaystyle \times \left[   \left(    \mathbf{\left[p \times  \left[\left(r-r' \right) \times p' \right]\, \right] \cdot }\frac{\partial}{\partial \mathbf{p}}  \right) \left(  \frac{f_{\alpha}(\mathbf{r},\mathbf{p},t)}{\sqrt{p^{2}+m_{\alpha}^{2}c^{2}} }\right)   \right]    = 0,}
\end{array}
\end{equation}
where $ \alpha=1,2,\ldots, M $, $  M $ is the number of types of particles contained in the system). This exact closed system of equations with excluded field variables describes the classical dynamics of a system of particles interacting through an electromagnetic field.

\section{Does a Hamiltonian of a system of interacting atoms exist?}
Since all substances consist of charged particles, the interaction energy of a system of atoms has an electromagnetic origin. Let us discuss the question of the possibility of representing the interaction energy of atoms through interatomic potentials, which depend on the simultaneous coordinates of all particles in the system.

\subsection{Electromagnetic energy of a system of resting charges}
The expression for the energy of the static electric field of resting charges has the form~\cite{Landau}:
\begin{equation}\label{static}
\begin{array}{c}
{\displaystyle  U = \frac{1}{8\pi} \int \left[ \sum_{s}\mathbf{E}_{s}\left(\mathbf{r} \right) \right]^{2} d\mathbf{r} = \frac{1}{8\pi} \sum_{s} \int \left[ \mathbf{E}_{s}\left(\mathbf{r} \right) \right] ^{2} d\mathbf{r}  }\\
{\displaystyle + \frac{2}{8\pi} \sum_{s<s'} \int \left(\mathbf{E}_{s}\left(\mathbf{r} \right) \cdot \mathbf{E}_{s'}\left(\mathbf{r} \right) \right) d\mathbf{r} . }
\end{array}
\end{equation}
The first term on the right side of this equation is the sum of the energies of the static fields of all particles. This energy is independent of particle positions.

The second term is the Coulomb energy of interactions between particles:
\begin{equation}\label{Coulomb}
\frac{2}{8\pi} \sum_{s<s'} \int \left(\mathbf{E}_{s} \left(\mathbf{r} \right) \cdot \mathbf{E}_{s'}\left(\mathbf{r} \right) \right) d\mathbf{r} = \sum_{s<s'} \frac{q_{s}\, q_{s'}}{\left| \mathbf{R}_{s} - \mathbf{R}_{s'}  \right|}.
\end{equation}

This energy is that {part of the energy of a static field}, which depends on the positions of particles in space~$ \mathbf{R}_{s} $. Thus, the Coulomb energy of interactions of resting charges is a part of the total energy of a static field, which depends on the distribution of these charges in space.

\subsection{Electromagnetic energy of a system of moving particles}

The energy conservation law for the classical system of charges in the general case has the following form~\cite{Landau}:
\begin{equation}\label{energy}
\int \frac{\mathbf{E}^{2}\left( \mathbf{r}, t\right) +\mathbf{H}^{2}\left( \mathbf{r}, t\right)}{8\pi}\ dV + \sum\mathcal{E}_{\mathrm{kin}}\left( t\right)  = \mathrm{const}. 
\end{equation}  
The first term in this formula is the energy of the electromagnetic field of the system, the second term is the sum of the kinetic energies of all particles.
Instantaneous value of electromagnetic field energy 
\begin{equation}\label{EM-energy}
\mathcal{E}_{\mathrm{em}}\left( t\right)  = \int \frac{\mathbf{E}^{2}\left( \mathbf{r}, t\right) +\mathbf{H}^{2}\left( \mathbf{r}, t\right)}{8\pi}\ dV
\end{equation}
is the functional of the scalar $ \varphi \left(\mathbf{r},t \right) $ and the vector $ \mathbf{A} \left(\mathbf{r},t \right)  $ potentials of the electromagnetic field at the same moment in time~$ t $. However, the potentials of the electromagnetic field at time $ t $ in accordance with the formulas~\eqref{varphi(r,t)} and~\eqref{A(r,t)} depend on the positions of the charges and the distribution of currents in space in all earlier time instants~$ t'\le t $.

Therefore, the interaction between atoms cannot be represented as an expression depending on the {simultaneous} positions $ \mathbf{R}_{s}\left(t \right)  $ of all particles of the system.

\section{Self-contradictoriness of statistical mechanics}
Thus, the potential energy of a system of moving interacting particles (in particular, atoms) does not exist. Consequently, the Hamiltonian, depending on the simultaneous coordinates and momenta of the particles of the system, does not exist too.

Note that the system of equations of motion for microscopic distribution functions, taking into account their construction, is relativistic. In this regard, it is appropriate to note that in the 1960s it was proved that the relativistic Hamiltonian formalism leads to the absence of interaction between particles~\cite{Currie,Currie2,Cannon,vanDam1,vanDam2}. Thus, the relativistically invariant Hamiltonian of the system of interacting particles does not exist, and the system of equations~\eqref{bas-eq} is non-Hamiltonian.

At the same time, the assumption of the existence of the Hamiltonian of a system of particles is one of the cornerstones of the construction of statistical mechanics. Taking into account the non-existence of the Hamiltonian, it is impossible to correctly determine neither the available volume of the phase space in the micro-canonical distribution, or partition functions in the canonical distribution.

Thus, the system of equations~\eqref{bas-eq}, which describes the dynamics of a system of particles with excluded field variables, is a non-Hamiltonian system. Therefore:
\begin{enumerate}
	\item the Liouville theorem on the conservation of phase volume, the Poincar\'{e} recurrence theorem, and the Liouville equation for phase density, which are proved for Hamiltonian systems, do not valid for a system of charges interacting through an electromagnetic field;
	
	\item the evolution equations of the microscopic distribution functions~\eqref{bas-eq} are not invariant with respect to the time reversal operation $ t \rightarrow -t $.
\end{enumerate}

Thus, the well-known Loschmidt and Zermelo paradoxes related to the dynamics of Hamiltonian systems are deduced within the framework of the classical dynamics of a system of charged particles interacting through an electromagnetic field: the exact system of equations~\eqref{bas-eq} is irreversible and the Poincar\'{e} recurrence theorem does not valid for this system of equations. In addition, the classical dynamics of a system of charged particles with excluded field variables can be put into the basis for consistent microscopic probability-free foundation (argumentation) of both physical kinetics and thermodynamics.

\section{Discussion}

The complete system of equations of dynamics of a closed classical (non-quantum) system of charged particles and the Maxwell equations of the electromagnetic field generated by them are transformed into an exact complete system of equations for microscopic distribution functions of particles by eliminating field variables.

The resulting system of functional equations, in addition to the fact that it does not contain field variables, has the following characteristic features.

\begin{enumerate}
	\item This system of equations is initially relativistic. In this regard, it is appropriate to mention that shortly after the publication of the fundamental work of Maxwell~\cite{Maxwell1}, the works of Lorentz~\cite{Lorenz} and Riemann~\cite {Riemann} were published, which formulated the idea of the retarded nature of electromagnetic interactions. Thus, Maxwell's electrodynamics is actually one of the sources of the theory of relativity, which is reflected in the title of Einstein's work~\cite{Einstein}(On the electrodynamics of moving bodies).
	After the creation of the special theory of relativity, the idea that the reason for the irreversibility is the retardation in interactions was expressed by Ritz~\cite{Ritz1}, and a little later a short joint work by Ritz and Einstein~\cite{Ritz2} was published, the authors of which spoke mutually diametrically opposite opinions about the origin of irreversibility. Ritz believed that irreversibility is due to a retardation in interactions, and Einstein believed that the nature of irreversibility has a probabilistic origin.
	The further development of kinetic theory was based on the concept of probability both in the non-relativistic approximation and in relativistic models. The construction of the relativistic kinetic theory was initiated by the work of J\"{u}ttner~\cite{Juttner1, Juttner2}, Tetrode~\cite{Tetrode}, Fokker~\cite{Fokker}, Synge~\cite {Synge}, Chernikov~\cite{Chernikov}, Kuzmenkov~\cite{Kuzmenkov1} et al. Currently, the relativistic kinetic theory is an intensively developing field with various applications in plasma physics, transport processes at different scales from nuclear matter to cosmology~\cite{Groot2,Cerc1,Liboff,Hakim,Veresch}. For all the variety of objects and methods of research within the framework of the classical relativistic kinetic theory, most modern works are united by two common ideas.
	\begin{itemize}
		\item (i) Initially, the kinetics is described in terms of not microscopic, but averaged (or smoothed) distribution functions.
		\item (ii) One of the mechanisms of irreversible behavior of a many-body system is collisions between particles associated with collision integrals. However, the collision integrals are expressed in terms of inter-particle potentials, which do not exist in relativistic classical dynamics. Therefore, the use of the collision integral, as well as the concept of probability, as a cause of irreversibility in the framework of the relativistic kinetic theory seems doubtful.
		\end{itemize}
		
	A variant of constructing the kinetic theory of a system of particles interacting through an arbitrary two-particle potential with account of retardation, in terms of a microscopic distribution function, was proposed in papers~\cite{Zakh3,Zakh4}. It is shown that the use of probabilistic concepts is not a necessary condition for for irreversible behavior of the system.
	
	Investigations of few-body systems, taking into account the retardation in interactions, are limited by the simplest particular case — the two-body problem. However, even this ``simplest task'' turned out to be extremely complex and dealt mainly with the two-body problem in electrodynamics~\cite{Synge1, Driver0, Driver}. In these works, it was shown that the retardation in interactions leads to a radical qualitative change in the dynamics of systems, including irreversibility with respect to the time reversal operation $ t \rightarrow -t $. This behavior, however, was not interpreted in these works as a possible mechanism of thermodynamic irreversibility. In papers~\cite{Lucia1,Lucia2} a mechanism of macroscopic irreversibility was developed as a consequence of \textit{microscopic} irreversibility: atoms are considered as open systems that continuously interact with environmental photons.
		
	In~\cite{Zakh6},  a model of a one-dimensional two-particle harmonic oscillator with a retarded interaction between the particles was studied. It is shown that the characteristic equation has infinitely many \textit{complex} roots. This means that in this system there are infinitely many disappearing and infinitely many incipient and increasing in time degrees of freedom. This was interpreted as a sign of irreversible, i.e. thermodynamic behavior of a two-particle oscillator. Thus, the irreversibility of the dynamics of a system of particles is a common property of both many-body and few-body systems, and this property is due to the finiteness of the propagation speed  of interactions between particles.
	\item The system of equations~\eqref{bas-eq} connects the functions  $ f_{\alpha}\left(\mathbf{r,p},t \right)  $ at instant of time $t $ with all the functions $ f_{\gamma}\left(\mathbf{r',p'},t' \right) $ at other points  $ \mathbf{r', p'} $ at all earlier instants of times $ t' \le t $. Therefore, the equations~\eqref{bas-eq} take into account both the property of spatial non-locality of the system and its heredity, that is, dependence of the evolution of a system of particles on its prehistory. The foundations of the hereditary model in the theory of elasticity were laid by Boltzmann~\cite{Boltzmann}, and the basic principles and some applications of the mathematical theory of heredity were developed by Volterra~\cite{Volterra-1, Volterra-2}.	

	\item The system of equations~\eqref{bas-eq} is neither a Hamiltonian nor a conservative system; therefore, it does not obey neither the Poincar\'{e} recurrence theorem, the Liouville theorem on conservation of phase volume, nor the Liouville equation for the evolution of phase density. The dynamics of the particle system in the general case is not quasi-periodic, and, in particular, the well-known Zermelo paradox does not arise.

	\item The system of equations~\eqref{bas-eq} is not invariant with respect to the time reversal operation $ t \rightarrow -t $, therefore the well-known Loschmidt paradox does not arise. It is essential that the irreversibility of the dynamics of a system of particles is not the prerogative of many-particle systems: the property of irreversibility takes place, generally speaking, also for few-body systems. We note that the irreversibility of the equations of motion of a system of particles is only a \textit{necessary} condition for the thermodynamic behavior of the system, but, generally speaking, \textit{not sufficient}, since \textit {establishment} of thermodynamic equilibrium in the system does not mean stopping the evolution of the system on scale of microscopic scale sizes.
\end{enumerate}

\section{Prospects}

Thus, a complete and exact system of equations of evolution of the classical system of charges~\eqref{bas-eq} is a consistent description of the dynamics of the system. The characteristic features of this system of equations are as follows.
\begin{itemize}
	\item It does not contain any additional hypotheses, including probabilistic assumptions such as the molecular chaos hypothesis, is free from internal contradictions such as the Loschmidt and Zermelo paradoxes, and does not require the use of (man-made) interatomic potentials.

	\item The system of equations~\eqref{bas-eq} satisfies all the conditions that are necessary for the microscopic justification of both relativistic and non-relativistic kinetic theory and thermodynamics. In particular, it is not invariant with respect to the time reversal operation, i.e. irreversible.
	
	\item  The usual setting of initial conditions such as Cauchy problems for finding solutions to this system of equations is incorrect due to the non-locality of the equations. In contrast to problems of classical (non-relativistic) mechanics, in which setting of the initial conditions results to an unambiguous \textit{deterministic} solution (Laplace determinism), solving of the system of equations~\eqref {bas-eq} requires knowledge of all functions  $ f_{\alpha}\left(\mathbf{r',p'},t' \right)  $ for all $ t' \leq t_ {0} $ ($ t_{0} $~is initial point in time). Note that all microscopic distribution functions for $ t '\leq t_ {0} $ cannot be given arbitrarily, since even for $ t' \leq t_ {0} $ these functions are predetermined by the same system of equations~\eqref{bas-eq}, but with earlier time instants $ t $. Therefore, unlike the Laplacean picture of the world, the values of microscopic functions at any instant of time depend on the \textit {full} prehistory of the system, including arbitrarily any amount remote moments in the past. On the other hand, this property of solutions for the system of equations~\eqref{bas-eq} means the universality of the heredity phenomenon, which, in particular, manifests itself in hysteresis phenomena.
	\item It should be pointed out on the extremely poor development of the mathematical apparatus of functional differential equations~\cite{Cheng}. Therefore, it is difficult to hope for the possibility of obtaining of exact solutions to even the simplest problems of the relativistic dynamics of charged particles. The most relevant are two variants for studying of the system of equations~\eqref{bas-eq}.
	The closest prospects are the search for methods of a qualitative analysis of the properties of solutions of the system of equations~\eqref{bas-eq} and the search for solutions in the form of expansion in powers of $ v / c $, where $ v $ are the characteristic velocities of the particles of the system. In particular, in the case of low particle velocities, it is easy to obtain the collisionless Boltzmann equation, Vlasov equations, and other approximations from the system of equations~\eqref{bas-eq}.
\end{itemize}

\section{Conclusion}
I am grateful to Profs.~Y.I.~Granovsky, V.V.~Uchaykin, and Dr.~V.V.~Zubkov for intensive discussions and constructive criticism.

This work was partially supported by the Ministry of Science and Higher Education of Russia within the framework of the project part of the State Assignment (grant No.~3.3572.2017, 2017-2019).

\end{document}